\documentclass[aps,prl,twocolumn,showpacs,amsmath,amssymb,superscriptaddress,floatfix]{revtex4}
\usepackage{graphicx,epsf}
\usepackage{bm}

\begin{document}

\title{Third edge for a graphene nanoribbon: A tight-binding model calculation.}

\author{D. A. Bahamon}
\address{Instituto de F\'\i sica, Universidade Estadual de Campinas - UNICAMP,  C.P. 6165, 13083-970, Campinas, Brazil}
\author{A. L. C. Pereira}
\address{Faculdade de Ci\^encias Aplicadas, Universidade Estadual de Campinas, Limeira, SP, Brazil}
\author{P. A. Schulz}
\address{Instituto de F\'\i sica, Universidade Estadual de Campinas - UNICAMP,  C.P. 6165, 13083-970, Campinas, Brazil}

\date{\today}

\begin{abstract}

The electronic and transport properties of an extended linear defect embedded in a zigzag nanoribbon of realistic width are studied, within a tight binding model approach. Our results suggest that such defect profoundly modify the properties of the nanoribbon, introducing new conductance quantization values and  modifying the conductance quantization thresholds. The linear defect along the nanoribbon behaves as an effective third edge of the system, which shows a metallic behavior, giving rise to new conduction pathways that could be used in nanoscale circuitry as a quantum wire.

\end{abstract}

\pacs{73.23.-b, 73.63.-b, 81.05.ue}


\maketitle


\begin{figure}[t]
\vspace{-0.2cm}
\centerline{\includegraphics[width=5.7cm]{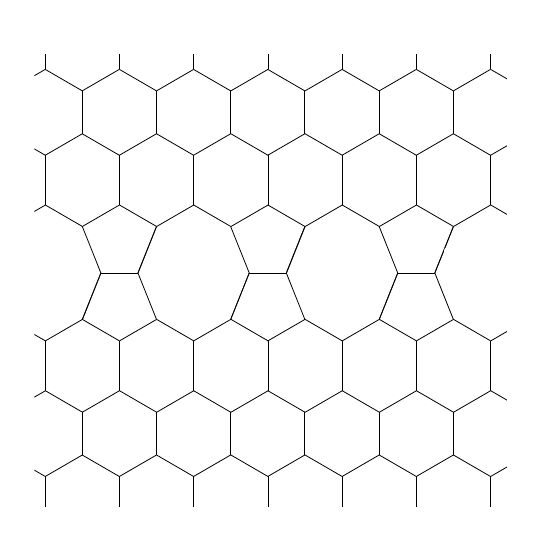}}
\vspace{-0.2cm}
\caption{ Geometry of the extended defect considered.}
\end{figure}

\section{I. Introduction}

Graphene, a unique purely two dimensional system, evolved already into a landmark in condensed mater physics, since its first successful isolation \cite{Novoselov1} and the initial establishment of their related electronic properties\cite{rev_mod_phys}. More recently, it has been addressed as to initiate a new generation of electronic devices \cite{Schwierz}, since the obtention of graphene nanoribbons \cite{nanoscience} and the cutting out of more complex structures by means of atomic force microscopy nanolitography \cite{Tseng,cutspseudocuts}.Nevertheless, many  applications require control at atomic scale \cite{Nakada}, which means that a conflicting scenario has to be still overcome. On one side, for instance, clear single electron charging has been observed in graphene quantum dots \cite{Ensslin}, but on the other side, there seem to be limitations to the use of graphene as ballistic nanowires due to still unavoidable disorder effects \cite{Mucciolo}. In this context, the introduction of localized defects has been suggested as one accessible way to modify the electronic properties of graphene \cite{Costa_Filho,Ferreira,Yazyev2}. Alternatively, the perspective of inducing extended defects, in particular linear defects, that could even assemble involved crossed structures over a large area on graphene, started to be discussed as a candidate for monolithic circuitry on a nanoscale \cite{Appelhans,defenglusk}. Besides first principle studies on the stability of such structures and the associated density of states, systematic evaluation of their related transport properties have started to be undertaken \cite{Yazyev}.

A promising achievement is the observation of an extended linear defect \cite{nw_defecto} on a  graphene layer grown on a metallic substrate. This defect is created by alternating Stone-Thrower-Wales (STW) defect \cite{Stone,Thrower} and di-vacancies, leading to a pattern of repeating paired pentagons and octagons, as shown in Fig 1. It has been shown, combining first principle calculations and experimental scanning probe microscopy observations, that this defect acts as a one dimensional metallic wire \cite{nw_defecto,Appelhans}, widening a road towards a graphene based electronics. 

Here we focus on a tight binding model calculation to scrutinize the conductance of such a linear defect embroidered along a graphene nanoribbon with outer zigzag edges. After validating the model by ensuring that the main features of the electronic properties of the linear defect are captured by a nearest neighbor tight binding Hamiltonian, the conductance is obtained within a Landauer Buttiker approach. The complex structure of the defect leads to sharp anti-resonances in the lower conductance plateaus, which, on the other hand, may either occur at even multiples of $G_0=\frac{2e^2}{h}$, rather unexpected for graphene systems or may define new conductance plateau thresholds, acting effectively as a third edge in a zigzag nanoribbon. The present results are relevant for the experimental context in which the controlled fabrication of extended defects \cite{ocomtrema} may be conbined with the nanolitographyc definition of nanoribbons \cite{cutspseudocuts}.

In section II we briefly sketch our model approach and proceed by validating it in respect to previous first principle calculations \cite{nw_defecto,Appelhans}. In section III the electronic band structure of an infinite zizgag nanoribbon with and embedded linear defect is shown. In section IV the conductance , as well as the real space current density distribution modification due to the presence of a linear defect is discussed, while in section V the final comments and experimental perpsectives are placed.

\section{II. Heuristic model for an extended defect}

We model the electronic structure by a nearest neighbor tight-binding Hamiltonian for an hexagonal lattice:

\begin{equation}
H = t  \sum_{<i,j>} (c_{i}^{\dagger} c_{j} +
c_{j}^{\dagger} c_{i})
\end{equation}

\hspace{-\parindent}where $c_{i}$ is the fermionic operator on site $i$. We model the planar arrangement of the atoms in the extended linear defect as shown in Fig. 1 by rearranging which atoms are connected by nearest-neighbor hopping parameters ($t$). This type of defect shows atomic separations ranging  from 1.38 {\AA} to 1.44 {\AA} and conserves the coordination number \cite{nw_defecto}, which implies variations lesser than 5$\%$ in $t$, suggesting that $t$ values could be considered nearly unaffected in respect to a defect free graphene nanoribbon (for which $t\approx 2.7$ eV). In order to compare the results from our model to previous calculations \cite{nw_defecto,Appelhans}, we calculate the density of states (DOS) for an infinite graphene sheet (18200 atoms in the unit cell) with the extended defect. The result is shown in Fig. 2 where a peak around the Dirac point (E/t =0) is developed, indicating the metallic character and the one dimensional nature of the defect at this energy, in accordance  to the already mentioned first principles calculations \cite{nw_defecto,Appelhans}.
The conductance $G(E)$ is evaluated within the Landauer-B$\ddot{u}$ttiker formalism, $G(E)=G_0T(E)$, where $G_0=\frac{2e^2}{h}$ is the conductance quantum  and $T(E)$ is the transmission function between the contacts and is evaluated by \cite{Datta}:

\begin{equation}
T_{pq}=Tr[\Gamma_pG^r_{pq}\Gamma_qG^\dagger_{pq}]
\end{equation}

\hspace{-\parindent} where  $G^r_{pq}$ is the Green function between the contact $p$ and $q$ evaluated thorough the recursive lattice Green's function technique \cite{PALee,Ferry}. The contact broadening function \cite{cresti1} $\Gamma_{p(q)}=i(\Sigma_{p(q)}-\Sigma^\dagger_{p(q)})$, where $\Sigma_{p(q)}$ is the self-energy of the contact, arises from the interaction of the central region with the semi infinite contacts. In order to calculate this term, it is necessary to know the Green function of the contact, also obtained numerically \cite{LSancho}. 

\begin{figure}[t]
\vspace{-0.2cm}
\centerline{\includegraphics[width=5.8cm]{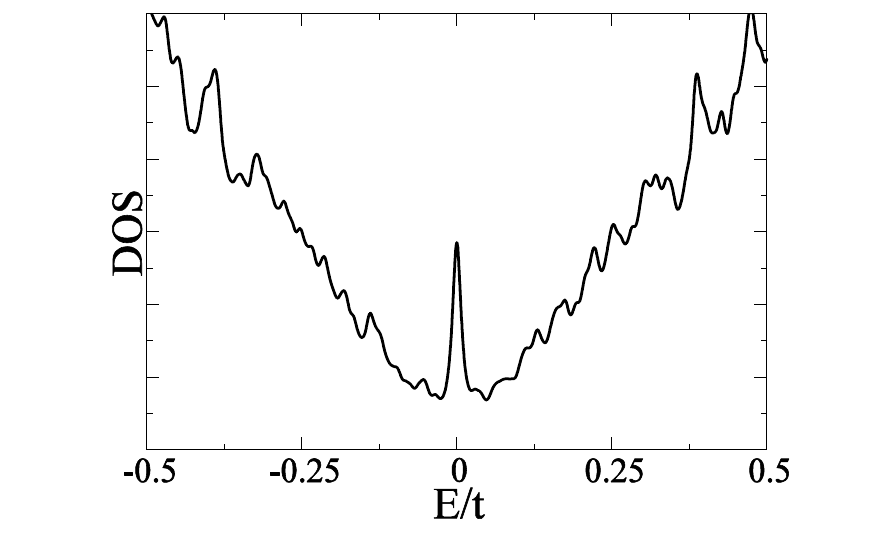}}
\vspace{-0.2cm}
\caption{Density of states (DOS) for an extended defect in an infinite graphene sheet. }
\end{figure}
	
\section{III. A wire embedded in a zigzag nanoribbon}

\begin{figure}[t]
\vspace{-0.2cm}
\centerline{\includegraphics[width=8.5cm]{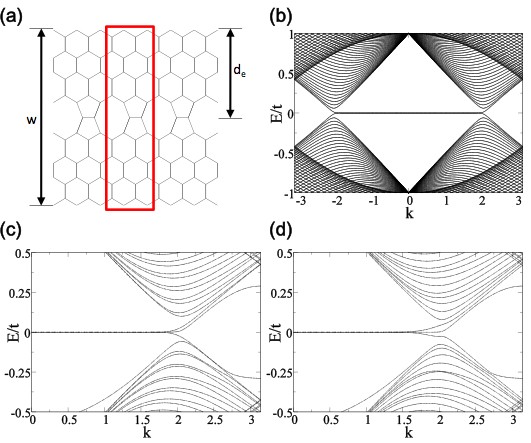}}
\vspace{-0.2cm}
\caption{(Color online)  {\bf (a)} Geometry of zigzag graphene nanoribbon of width $W$ with an extended defect at a distance $d_e$ from the upper edge.  {\bf (b)} Band structure of a zigzag nanoribbon of width $W=15.9$ nm without any defect. {\bf (c)} Band structure of a zigzag nanoribbon of $W=15.9$ nm with an extended defect  at $d_e=8$ nm .{\bf (d)} Band structure of a zigzag nanoribbon of $W=15.9$ with an extended defect  at $d_e=3.8$ nm  }
\end{figure}

The band structure of an infinite linear defect embedded along a zigzag graphene nanoribbon is obtained within the scheme illustrated in Fig. 3(a), where the unit cell used for the calculation is also shown. The objectives here are twofold: on one hand we have to compare the band structure obtained within our model and for a nanoribbon with the band structure of a bulk infinite graphene sheet, as presented previously \cite{nw_defecto}. On the other hand, we want to capture the changes in the band structure induced by the linear defect looking forward the interpretation of the conductance features of a scattering region of finite length (where the defect is embedded) terminated by semi infinite zigzag nanoribbon contacts of the same width.  The band structure of a zigzag nanoribbon $W=15.9$ nm wide, without the extended defect, is depicted in Fig. 3(b): a flat band at zero energy, corresponding to edge states, is developed  in the range $-2\pi/3 < k < 2\pi/3$. Notice that usually these flat bands are seen within $2\pi/3 < |k| < \pi$ \cite{Nakada}. Our slightly different representation is due to a folding introduced by the fact that we use a unit cell twice as large as necessary for a defect less nanoribbon, in order to properly describe the defect afterwards. The presence of the extended defect in the nanoribbon does not alter the metallic character or the appearance of the edge states in the nanoribbon, but clearly (i) breaks the particle-hole symmetry \cite{pereirabreak}  in a new context, due to the rebounding along the extended defect  and (ii) creates a new flat band at zero energy (now in a range around $k=0$, in accordance to what has been obtained for the bulk case \cite{nw_defecto}) which bends upwards in energy as shown in Fig. 3(c),  for an extended defect  located in the middle ($d_e=8$ nm) of the nanoribbon. Mapping the charge density (not shown here)  for different values of $k$ related to this band, one could see that the wave function spreads out mainly along the edges with a significant contribution at the defect sites, a contribution that rapidly decays away from the defect region. This decay of the wave function away from the defect has been already measured \cite{nw_defecto} and our findings within the present heuristic model are qualitatively in accordance with the experimental data. The degeneracy of the states at zero energy is broken at different $k$ values depending on the width of the nanoribbon or by simply putting the defect closer to one edge, the case shown in Fig. 3(d), depicting the band structure when the extended defect is located at $d_e=3.8$ nm, i.e., roughly halfway towards an edge from the middle, in the same nanoribbon of $W=15.9$ nm. Comparing Figs. 3(c) and 3(d), the stronger warping of the defect-like band is stronger for a defect line closer to the edge, but the effect seems to be qualitatively the same, irrespective to the distance to one of the outer edges. More importantly is that, comparing the defect free nanoribbon, Fig. 3(b), with Fig. 3(c) and 3(d), one sees the clustering in pairs of the bulk bands, a fingerprint of a division of the original nanoribbon into an effective double ribbon in parallel. These results for an infinite system constitute key ingredients to understand the new conductance features.

\section{IV. Conductance in the presence of the extended defect}

\subsection{A. Modified conductance quantization}

The conductance is calculated for nanorribons of different lengths L, containing longitudinal linear defects. Recalling, for sake of clearness, these systems are attached to two semi-infinite contacts of ideal graphene zigzag nanoribbons of the same width $15.9$ nm. The conductance of a perfect nanoribbon shows quantized conductance plateaus following an odd sequence specific rule  \cite{Lin,Peres}:

\begin{equation}
 G(E)=G_0(2n+1)
 \end{equation}

\hspace{-\parindent} while the conductance plateaus onset energies are given by \cite{Tworzydlo}:

\begin{equation}
\frac{E_n}{|t|}=\pm (n +\frac{1}{2})\frac{\sqrt{3}a\pi}{2 W}
 \end{equation}

\hspace{-\parindent} where $G_0=\frac{2e^2}{h}$, $n$ indicates the number of the conducting channel, $a=2.4$ $\AA$  is the graphene lattice constant and $W$ is the nanoribbon width. For $W=15.9$ nm $\frac{E_1}{|t|}\approx \pm 0.06$ and  $\frac{E_2}{|t|}\approx \pm 0.11$. The ideal conductance line shapes for defect free nanoribbon for the present parameters are plotted in Fig. 4 [(a)-(d)] as dashed lines. However, in Figs. 4(e) and (f), the dotted lines are for the conductance of a nanoribbon with an {\it  infinite} linear defect, corresponding to the band structure shown in Fig. 3 (c) and ((d), respectively. The actual conductance, i.e., including the {\it finite} longitudinal defects, is depicted as continuous lines. In what follows we discuss the important and also departures from the transport of an ideal zigzag nanoribbon. First of all, Fig. 4 shows nanoribbons of different length (with the linear length of the order of the nanoribbon), namely $L=50$ nm and $L=375$ nm, and different defect to edge distances: $d_e=8$ nm, i.e. in the middle of the ribbon, and $d_e=3.8$ nm, half way from the middle towards one of the edges.

Let us consider first Fig. 4(a), a shorter defect in the middle of the ribbon, as a starting point. Initially, one has to recall the band structures in Fig.3 and the particle-hole symmetry breaking, with one of flat bands bending downwards as in ideal ribbons, while a defect related band bends upwards together with a edge related band. Therefore the conductance shows two different kinds of oscillations: there are no qualitative changes in the first hole-like plateau region, the oscillations shown in this energy range are simply Fabry-Perot interference effects due to the broken translational symmetry introduced by the defect. On the other hand, the first electron-like plateau shows not interference oscillations but  strong anti-resonances \cite{wakabayashi}, due to the coupling of extended states at the edges with localized ones in the defects, as discussed elsewhere for a double vacancy system \cite{yo2}. Notice that these anti resonances change into Fabry-Perot oscillations in the second electron-like plateau. This is associated to a change from localized to extended character of the defect related states in this energy range, where, besides the interference oscillations, the conductance maxima reaches $2G_0$ instead of the expected $3G_0$ for clear graphene nanoribbons. 

\begin{figure}[t]
\vspace{-0.2cm}
\centerline{\includegraphics[width=8.5cm]{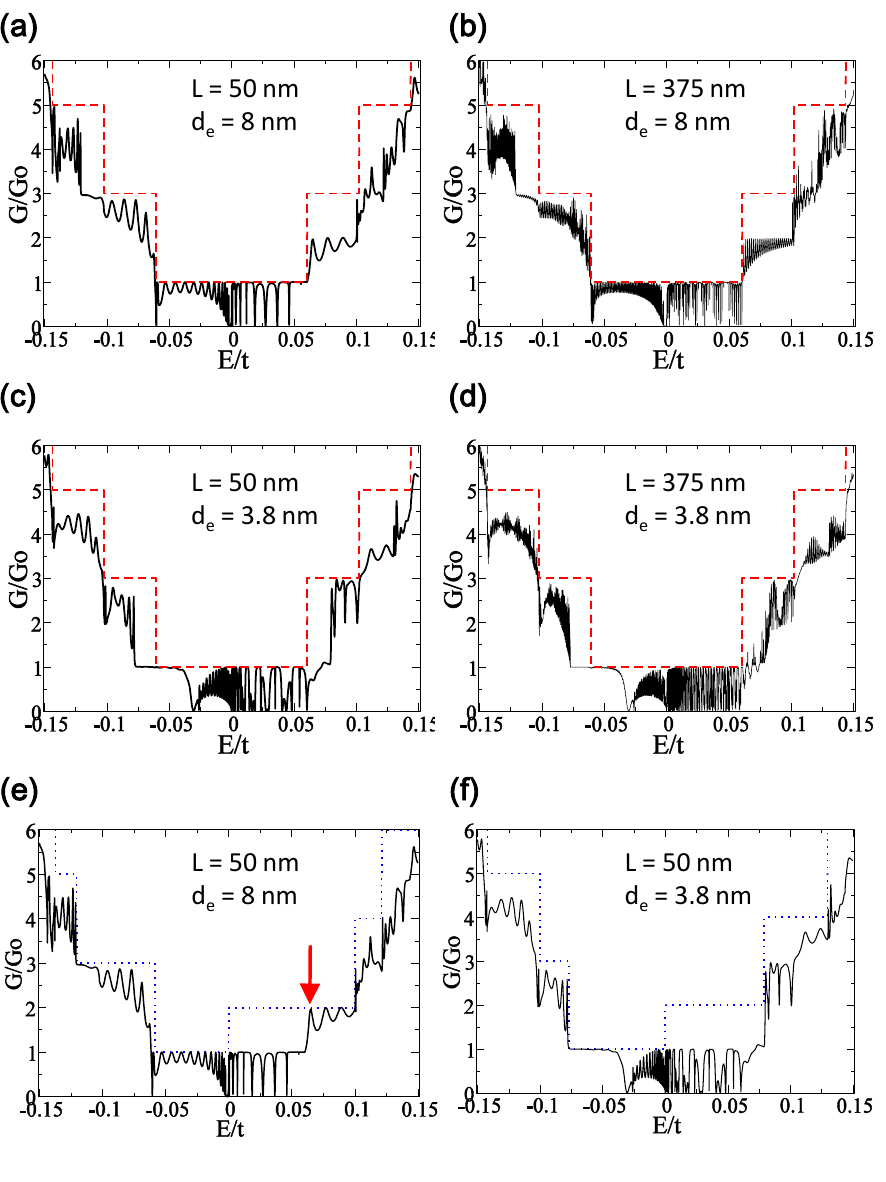}}
\vspace{-0.2cm}
\caption{ (Color Online)  Dashed lines correspond to the conductance for a perfect zigzag graphene nanoribbon of width $W=15.9$ nm. Solid lines correspond to the conductance for an extended defect inserted in a ribbon with {\bf (a)} ribbon width $W=15.9$ nm, defect length $L=50$ nm and distance from defect to the upper edge $d_e=8$ nm. {\bf (b)} $W=15.9$ nm, $L=375$ nm and $d_e=8$ nm. {\bf (c)} $W=15.9$ nm, $L=50$ nm and $d_e=3.8$ nm.  {\bf (d)} $W=15.9$ nm, $L=375$ nm and $d_e=3.8$ nm. {\bf (e)} Dotted line correspond to the conductance of a zigzag nanoribbon, $W=15.9$, with an infinite linear defect at $d_e=8$ nm. Continuous line $W=15.9$ nm, defect length $L=50$ nm  and $d_e=8$ nm. {\bf (f)} Dotted line correspond to the conductance of a zigzag nanoribbon, $W=15.9$, with an infinite linear defect at $d_e=3.8$ nm. Continuous line $W=15.9$ nm, defect length $L=50$ nm  and $d_e=3.8$ nm. Arrow refers to Fig. 5}
\end{figure}

Looking at a longer defect system, Fig. 4(b), one sees a clearer definition of the even electron conductance plateau, besides the increase in the number of interference oscillations, as well as anti resonances in all plateaus, simply due to increasing the length of the extended defect. Here one should pay attention to the third electron conductance plateau that evolves in a non-trivial way towards an expected $5G_0$ value. Besides the strong oscillations, analogous to the already discussed ones in lower plateaus, it can be seen that this plateau splits into two sub plateaus, one reaching $3G_0$ and only the second approaching $5G_0$. This plateau splitting, occurring also at other higher plateaus and in the hole side of the spectrum, is due to the band clustering in pairs, already discussed in respect to Fig. 3. Indeed, the onsets of these sub plateaus are given by the onset of subsequent bands in Fig. 3.

Moving now the defect towards one of the edges, new features appear, as observed in Figs. 4(c) and 4(d). The most striking one is the fact that the onset of second plateau onset, either for holes and electrons can not be associated to the corresponding  ideal nanoribbon, as happens for the linear defect in the middle. It is worth noting that these effective conductance plateau onsets correspond to a thinner nanoribbon, namely $W=12.1$ nm wide by fitting the numerical results in Fig. 4 by eq. (4). This corresponds exactly to the distance from the linear defect to the more distant edge: $15.9$ nm - $3.8$ nm $=$ $12.1$ nm, recalling that the width of the nanoribbon is $W=15.9$ nm and $d_e=3.8$ nm is the distance of the defect to one of the edges, for the results shown in Fig. 4(c) and Fig. 4(d). This behavior, potentially observable, is explained by the fact that for $d_e=3.8$ nm, the linear defect is strongly coupled to the nearest edge, defining, therefore a thinner effective nanoribbon.
Another striking effect is the wide anti resonance shown in the first hole-like plateau, which can also be directly associated to the band structure for the corresponding infinite system, Fig. 3(d): the  edge state band that bends downwards presents a second flat step at the energy of the anti-resonance.

To further interpret  our preceding findings about the conductance features due to the extended defect, a direct comparison to the conductanceof ribbons with an infinite linear defect, i.e., corresponding directly to the band structure in Fig.3, becomes necessary. These results are given as dotted lines in Figs. 4(e) and 4(f), considering both $d_e=8$ nm $d_e=3.8$ nm, respectively. 
Although some features, like the existence of a $2G_0$ electron-like plateau, are revealed, the finite length defect nanoribbon is indeed an involved system, that can not be completely understood by means of infinite counterparts. In particular, although the prediction of a $2G_0$ plateau already at the Dirac point, the actual conductance for low energy electrons reaches only $G_0$, as expected from a defect free nanoribbon, actually the description of the leads connected to the defect region. Hence, the finiteness of the linear defects introduces unexpectancies. The supression of the conductance  at the first electron-like plateau in Fig. 4(e) and 4(f) are due to backscattering of a defect related mode at the end of the connection to the leads. 

\subsection{B. Metallic third edge}

\begin{figure}[t]
\vspace{-0.2cm}
\centerline{\includegraphics[width=8.5cm]{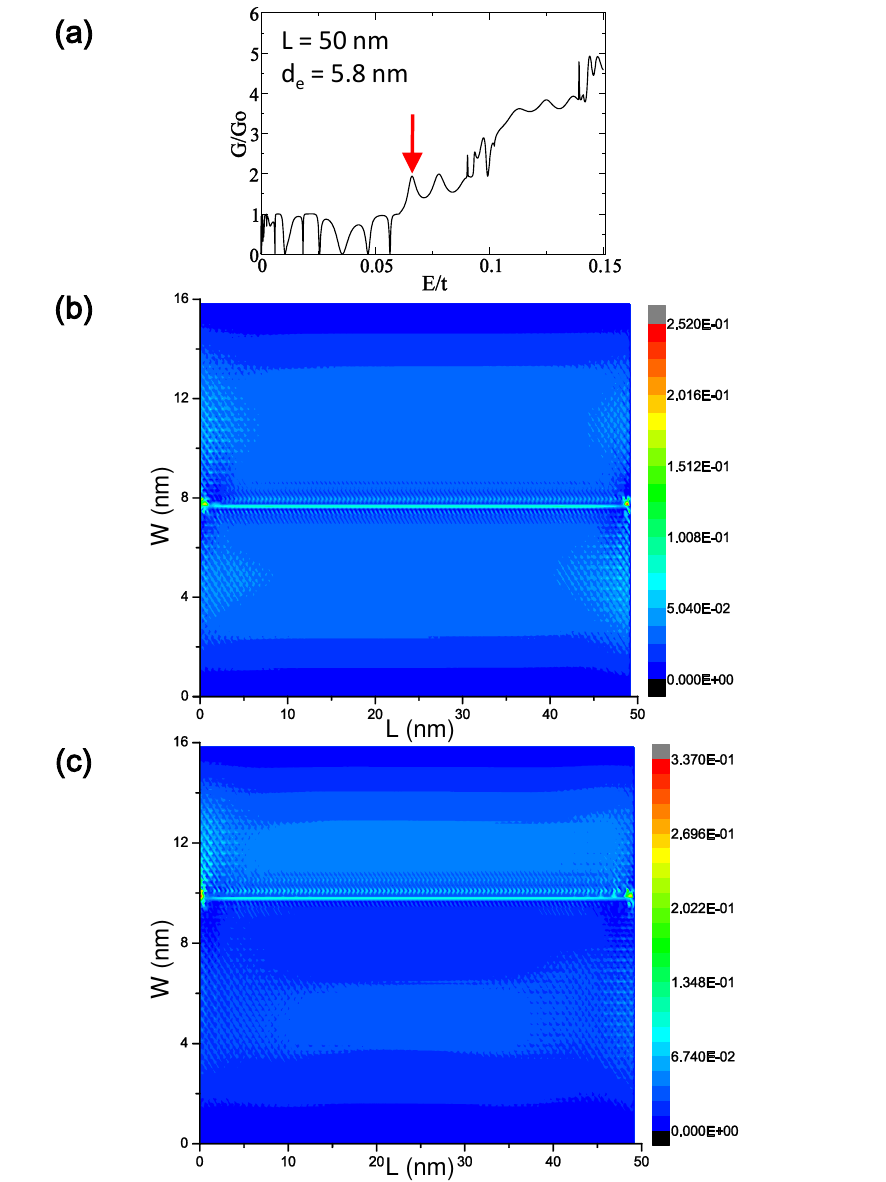}}
\vspace{-0.2cm}
\caption{ (Color Online)  {\bf (a)}  Conductance of a zigzag ribbon width $W=15.9$ nm with a linear defect of length $L=50$ nm and distance from defect to the upper edge $d_e=5.8$ nm. {\bf (b)} Real space current density distribution, in units of $G_0=\frac{2e^2}{h}$, for the conductance peak indicated by the arrow in Fig. 4(e).  {\bf (c)} Real space current density distribution, in units of $G_0$, for the conductance peak indicated by the arrow in Fig. 5(a). }
\end{figure}

To further investigate the linear defect characteristics, the real space current density distribution is depicted in units of $G_0$, following the procedure sketched in \cite{cresti1,cresti2}, Fig. 5. The arrow in Fig. 4(e) points to the first conductance peak reaching $3G_0$ at an energy ($E/|t|=0.0651$), associated to the real space current density distribution shown in Fig. 5(b). A high current density  around the defect reveals its metallic character, although the current spreads out considerably across the ribbon, differently from defect free system, where the low energy current density is located mainly at the center of the nanoribbon \cite{Zarbo}. A closer look shows that the defect acts as a third edge, since its presence creates a higher current density in the middle of the two newly defined nanoribbons around the linear defect. Interference fringes on both sides of this central edge, are also appreciated, as well as in the region in contact with the pristine graphene nanoribbon leads. A similar behavior is found (high current density located along the embeded linear structure) if the current density is plotted for the others conductance peaks in this electron plateau (not shown here). In Fig 5(a) the conductance of a nanoribbon of width $W=15.9$ nm, $d_e=5.8$ nm and $L=50$ nm, i.e., an off center embedded edge, is shown, where again the second electron plateau reaches the value of $2G_0$. The arrow points to the  first conductance  peak defined at an energy  of $E/|t|=0.0662$ and the real space current density distribution at this energy is shown in Fig 5(c).  The high current density along the defect is appreciated as well as its role as a third edge, defining two clearly different regions with different current densities, loosing now the symmetry of the current density between the upper ribbon and the lower ribbon.

\begin{figure}[t]
\vspace{-0.2cm}
\centerline{\includegraphics[width=8.5cm]{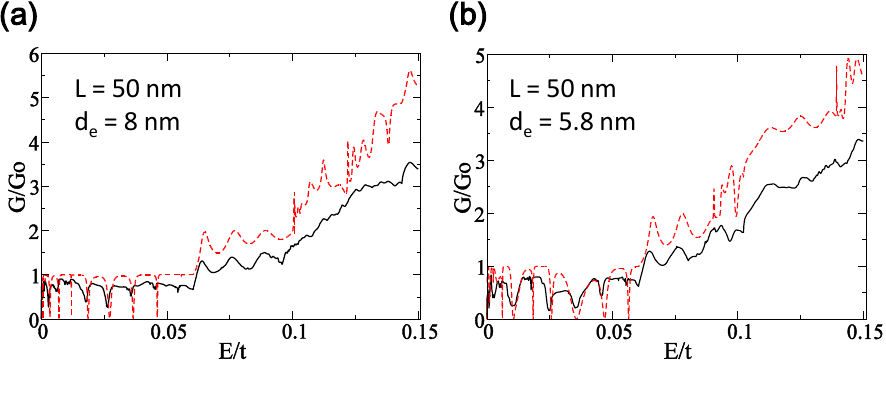}}
\vspace{-0.2cm}
\caption{ (Color Online)  {\bf (a)} Average Conductance of a zigzag ribbon of width $W=15.9$ nm, defect length $L=50$ nm and distance from defect to the upper edge $d_e=8$ nm with edge disorder. Dashed line without edge disorder.  {\bf (b)} Average Conductance of a zigzag ribbon of width $W=15.9$ nm, defect length $L=50$ nm and distance from defect to the upper edge $d_e=5.8$ nm with edge disorder. Dashed line without edge disorder. }
\end{figure}

One could envisage the defect as part of nanoscale circuitry, such system can be created by scanning probe microscopy patterning a nanoribbon \cite{Tseng} in an appropriate region of a graphene sheet containing such a linear structure. If the linear defect is a rather ordered structure \cite{nw_defecto}, the patterned outer edges  will become quite disordered. To study the effect of a disordered edge, lattice sites from both outer edges are removed \cite{Mucciolo,Ihnatsenka} with a probability $p_e=0.3$ and an average over twenty edge disorder realizations  was carried out. The average conductance for a nanoribbon of width $W=15.9$ nm, $d_e=8$ nm and $L=50$ nm is shown as the continuous line of Fig. 6(a). Comparing with the dashed line (same nanoribbon without outer edge disorder) we find that the anti-resonances of the first channel are robust structures, while the peaks higher channels are completely washed out, except for those three peaks in the second electron plateau. This robustness is associated, as already shown, by the fact that they have a high current density along the defect. In Fig. 6(b) we show, as a continuous line, the average conductance of the same nanoribbon, but moving the defect toward one edge, $d_e= 5.8$ nm, comparing with the dashed line (same nanoribbon without edge disorder). The main features are completely washed out except for the peaks which have a high current density along the defect. These results suggest that several conductance fingerprints, introduced by the linear defect, are robust against outer edge disorder showing that these extended defect structures could be used in real mesoscopic devices as a quantum wire. Such devices could be an interesting alternative to other potential modulation proposals \cite{Costa_Filho,Xie}. 

\section{V. Final comments}

In summary, we calculate the transport properties of a linear defect, recently observed in bulk graphene \cite{nw_defecto}, embedded in a zigzag nanoribbon.  Our model calculation, when compared to first principle results \cite{nw_defecto}, reveal that the main qualitative features have been captured. Since the bond lengths around the linear defect are modified, a further consistency check has been undertaken, namely modifying the hopping parameter $t$ according to the change in the bond lengths \cite{VPereira2}. Such procedure revealed no significant changes in our results \cite{Klos}. Looking therefore to the conductance, a variety of interesting fingerprints emerge. One of them is the presence of anti-resonances for electrons and bona fide Fabry-Perot oscillations for holes. Such features should be still difficult to observe considering present device quality. More robust fingerprints, having in mind possible experimental observation should be three fold: (i) presence of an even conductance plateau, (ii) a wide anti resonance in the first hole like plateau and, (iii) new positions of the conductance plateaus onsets, for the redefinition of the effective nanoribbon width by the linear defect, justifying the title of the present paper, that defines the linear defect as third edge for the nanoribbon. Indeed, the analysis of current density distribution and conductance calculations in presence of outer edge disorder suggest that many conductance fingerprints, related to appreciable current density around the linear defects, are robust against disorder which would be expected in real samples. Such samples could be obtained by means of ion beam irradiation \cite{ocomtrema} to create the extended defects, while the nanoribbon structure around the defect would be achieved by atomic force microscopy nanolitography \cite{Tseng}.

DAB acknowledges support from CAPES, ALCP acknowledges support from FAPESP. PAS received partial support from CNPq. Numerical calculations were developed at CENAPAD-SP and IFGW (UNICAMP) computing clusters.


\end{document}